\documentclass[prl,twocolumn]{revtex4}
\usepackage{graphicx}
\usepackage{amsfonts,amsmath,amssymb,float}
\usepackage{bm,dsfont}
\usepackage{color}
\usepackage{bbm}
\usepackage{longtable}
\usepackage[dvips]{epsfig}
\usepackage{amsmath,amssymb,lscape,float}
\usepackage{hyperref}
\usepackage{listings}

\begin{document}
\title{Bare-Excitation Ground State of a Spinless-Fermion -- Boson Model and 
$W$-State Engineering in an Array of Superconducting Qubits and Resonators}

\author{Vladimir M. Stojanovi\'c}
\email{vladimir.stojanovic@physik.tu-darmstadt.de}
\affiliation{Institut f\"{u}r Angewandte Physik, Technical
University of Darmstadt, D-64289 Darmstadt, Germany}

\date{\today}

\begin{abstract}
This Letter unravels an interesting property of a one-dimensional lattice model that describes a single itinerant spinless fermion (excitation)
coupled to zero-dimensional (dispersionless) bosons through two different nonlocal-coupling mechanisms. Namely, below a critical value of the 
effective excitation-boson coupling strength the exact ground state of this model is the zero-quasimomentum Bloch state of a bare (i.e., completely
undressed) excitation. It is demonstrated here how this last property of the lattice model under consideration can be exploited for a fast, 
deterministic preparation of multipartite $W$ states in a readily realizable system of inductively-coupled superconducting qubits and microwave 
resonators.
\end{abstract}

\maketitle

Sophisticated quantum-state engineering~\cite{ZagoskinBOOK,Makhlin+:01} is a prerequisite for the development of next-generation 
quantum technologies~\cite{Dowling+Milburn:03,Schleich+:16}. In this context, tantalizing progress was made in recent years by 
utilizing diverse physical platforms~\cite{Haas+:14,Friis+:18,Song+:17}. In particular, owing to their continuously improving scalability 
and coherence properties superconducting (SC) circuits~\cite{Xiang+:13,VoolDevoretReview:17,SCreviews:17,SCqubitReview:19} and, 
among them, circuit-QED systems~\cite{Koch++:07,GirvinCQEDintro}, allow accurate preparation of various quantum states of SC qubits 
and photons alike~\cite{Vlastakis+:13}.

The most prominent classes of entangled many-qubit states are maximally-entangled Greenberger-Horne-Zeilinger (GHZ)~\cite{Greenberger+Horne+Zeilinger:89} 
and $W$ states~\cite{Duer+:00}. An $N$-qubit $W$ state is the equal superposition of states with exactly one qubit in its ``up'' state, the 
remaining ones being in their ``down'' states. In particular, it is known that a $W$ state and its GHZ counterpart cannot be transformed into 
each other via local operations and classical communication (LOCC-inequivalence)~\cite{Nielsen:99}. $W$ states are also extremely robust with 
respect to particle loss, remaining entangled even if any $N-2$ parties lose the information about their particle~\cite{Koashi+:99}. They lend 
themselves to applications in quantum-information protocols~\cite{Joo+:03,Haeffner+:05,Haas+:14,McConnell+:15}, which motivates their preparation 
in various systems~\cite{Gangat+:13,Li+Song:15,Reiter+:16,Kang+:16,Chen+:17,Fang+:19,Souza+:19}.

This Letter establishes a connection between a one-dimensional (1D) lattice model describing a nonlocal interaction of a spinless-fermion 
excitation with dispersionless bosons and multipartite $W$ states. Its point of departure is the notion that in one of the relevant regimes 
this model -- which includes excitation-boson (e-b) couplings of Peierls and breathing-mode types -- has an unconventional ground state. 
Namely, below a critical value of the effective e-b coupling strength its ground state is the zero-quasimomentum Bloch state of a bare excitation. 
It is shown here how this property of the model under consideration can be exploited for a fast, deterministic preparation of $N$-qubit $W$ states 
in an array of inductively-coupled SC qubits and resonators, an analog simulator of this model~\cite{Stojanovic+Salom:19}.

The state-preparation protocol proposed here, based on microwave pumping, allows one to obtain multipartite 
$W$ states within time frames three orders of magnitudes shorter than the currently achievable coherence times of SC qubits. 
Unlike the situation in quantum-state control, where typical preparation times scale unfavorably with the system size, here 
they do not depend on the number of qubits at all. What makes this protocol particularly robust is the fact that its target 
state is the ground state of the system in a parametrically large window of values of its main experimental knob -- an external 
dc flux.

\textit{Model and its ground state}.-- The 1D lattice model under consideration describes a single spinless-fermion excitation 
interacting with dispersionless bosons through two different nonlocal coupling mechanisms. The noninteracting part of its total 
Hamiltonian includes the excitation kinetic-energy- and free-boson terms:
\begin{equation}\label{H_0}
H_{0} = -t_e\sum_n (c^\dagger_{n+1}c_n + \mathrm{H.c.}) 
+ \hbar\omega_{\textrm b}\sum_n b^\dagger_n b_n \:.
\end{equation}
Here $c^\dagger_n$ ($c_n$) creates (destroys) an excitation at site $n$ ($n=1,\ldots,N$), $b^\dagger_n$ 
($b_n$) a boson with frequency $\omega_{\textrm b}$ at the same site, while $t_e$ is the excitation hopping 
amplitude. The interacting (e-b) part is given by
\begin{eqnarray}\label{H_eb}
 H_{\textrm{e-b}} &=& g\hbar\omega_{\textrm b}\sum_n\Big[c^\dagger_n c_n
(b^\dagger_{n-1} + b_{n-1} - b^\dagger_{n+1} - b_{n+1}) \nonumber\\
&+&(c^\dagger_{n+1}c_n+\mathrm{H.c.})(b^\dagger_{n+1} + b_{n+1} 
-b^\dagger_n - b_n)\Big]\:,
\end{eqnarray}
where $g$ is the dimensionless e-b coupling strength. The first term on the right-hand-side (rhs) of the last equation captures 
the antisymmetric coupling of the excitation density at site $n$ with the local boson displacements on the neighboring sites $n\pm 1$ 
(breathing-mode-type coupling)~\cite{Slezak++:06}. The second term accounts for the linear dependence of the effective excitation-hopping 
amplitude between sites $n$ and $n+1$ on the respective boson displacements (Peierls-type coupling)~\cite{Stojanovic+:04,
Hannewald+:04,Stojanovic+Vanevic:08}.

\begin{figure}[t!]
\includegraphics[clip,width=8.3cm]{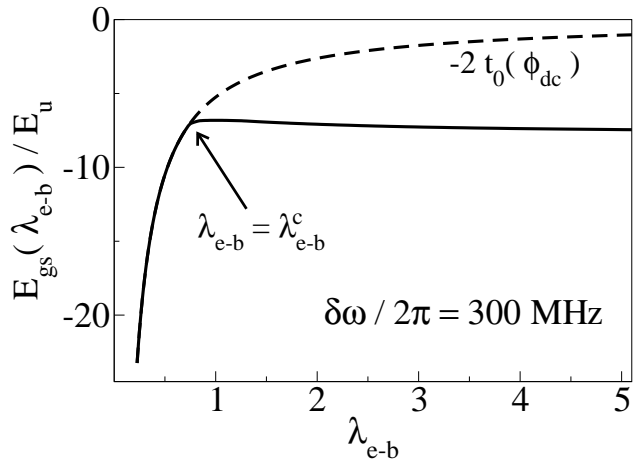}
\caption{\label{fig:Energy}Ground-state energy of the system with $\delta\omega/2\pi=300$\:MHz 
as a function of the effective coupling strength $\lambda_{\textrm{e-b}}$. For $\lambda_{\textrm{e-b}}
<\lambda^{\textrm{c}}_{\textrm{e-b}}\approx 0.72$ (i.e., $\phi_{\textrm{dc}}<0.972\:\pi$) the ground 
state of the system corresponds to a bare excitation, while for $\lambda_{\textrm{e-b}}\ge
\lambda^{\textrm{c}}_{\textrm{e-b}}$ it corresponds to a heavily-dressed (polaronic) one.}
\end{figure}

The coupling Hamiltonian $H_{\textrm{e-b}}$ can be recast in the generic momentum-space form 
\begin{equation}\label{Heb_ms}
H_{\textrm{e-b}}=\frac{1}{\sqrt{N}}\sum_{k,q}\gamma_{\textrm{e-b}}(k,q)\:
c_{k+q}^{\dagger}c_{k}(b_{-q}^{\dagger}+b_{q})\:. 
\end{equation}
Its corresponding e-b vertex function depends on both the excitation- and boson quasimomenta ($k$ and $q$, 
respectively, here expressed in units of the inverse lattice period) and is given by
\begin{equation}\label{vertex_func}
\gamma_{\textrm{e-b}}(k,q)=2ig\hbar\omega_{\textrm b}\:[\:\sin k+\sin q-\sin(k+q)] \:.
\end{equation}
The ground state of $H=H_0+H_{\textrm{e-b}}$ undergoes a sharp level-crossing transition~\cite{Stojanovic:20} at a 
critical value $\lambda^{\textrm{c}}_{\textrm{e-b}}\sim 1$ [cf. Fig.~\ref{fig:Energy}] of the effective coupling strength 
$\lambda_{\textrm{e-b}}\equiv 2g^{2}\:\hbar\omega_{\textrm b}/t_{e}$. For $\lambda_{\textrm{e-b}}<\lambda^{\textrm{c}}_{\textrm{e-b}}$ 
the ground state is the $K=0$ eigenvalue of the total quasimomentum operator $K_{\mathrm{tot}}=\sum_{k}k\:c_{k}^{\dagger}c_{k}
+\sum_{q}q\:b_{q}^{\dagger}b_{q}$. For $\lambda_{\textrm{e-b}}\ge\lambda^{\textrm{c}}_{\textrm{e-b}}$, on the other hand, 
the ground state is twofold-degenerate and corresponds to $K=\pm K_{\textrm{gs}}$ ($K_{\textrm{gs}}\neq 0$). 

A ground state with $K=0$ is by no means unusual -- in fact, an overwhelming majority of coupled e-b models 
have such ground states. Yet, the model at hand has the peculiar property that its $K=0$ ground state 
for $\lambda_{\textrm{e-b}}<\lambda^{\textrm{c}}_{\textrm{e-b}}$ is the $k=0$ bare-excitation Bloch state 
$|\Psi_{k=0}\rangle \equiv c^{\dagger}_{k=0}|0\rangle_{\textrm{e}}\otimes|0\rangle_{\textrm{b}}$, where 
$|0\rangle_{\textrm{e}}$ and $|0\rangle_{\textrm{b}}$ are the excitation and boson vacuum states.
In what follows, it will first be demonstrated explicitly that $|\Psi_{k=0}\rangle$ is an exact eigenstate 
of $H$ for an arbitrary value of $\lambda_{\textrm{e-b}}$. It will subsequently be shown numerically 
(see Fig.~\ref{fig:Energy} below) that for $\lambda_{\textrm{e-b}}<\lambda^{\textrm{c}}_{\textrm{e-b}}$ 
this state is the ground state of $H$. 

Given that $|\Psi_{k=0}\rangle$ is an eigenstate of $H_0$, to prove that it is an eigenstate of the total 
Hamiltonian $H$ it suffices to show that it is also an eigenstate of $H_{\textrm{e-b}}$. Indeed, by acting 
with $H_{\textrm{e-b}}$ [cf. Eq.~\eqref{Heb_ms}] on this state and making use of the fact that 
$c_{k}c_{0}^{\dagger}\:|0\rangle_{\textrm{e}}\equiv\delta_{k,0}|0\rangle_{\textrm{e}}$, 
one obtains
\begin{equation}\label{vanish}
H_{\textrm{e-b}}|\Psi_{k=0}\rangle = \frac{1}{\sqrt{N}}\sum_{q}\gamma_{\textrm{e-b}}(k=0,q)\:
c_{q}^{\dagger}|0\rangle_{\textrm{e}}\otimes b_{-q}^{\dagger}|0\rangle_{\textrm{b}}\:.
\end{equation}
Because here $\gamma_{\textrm{e-b}}(k=0,q)=0$ for an arbitrary $q$ [cf. Eq.~\eqref{vertex_func}], each term in the sum on 
the rhs of Eq.~\eqref{vanish} vanishes, implying that $H_{\textrm{e-b}}|\Psi_{k=0}\rangle=0$. Therefore, $|\Psi_{k=0}\rangle$ 
is an eigenstate of $H_{\textrm{e-b}}$ (for an arbitrary $\lambda_{\textrm{e-b}}$), the corresponding eigenvalue 
being equal to zero. This concludes the proof that $|\Psi_{k=0}\rangle$ is an exact eigenstate of $H$. 

\begin{figure}[b!]
\includegraphics[clip,width=8cm]{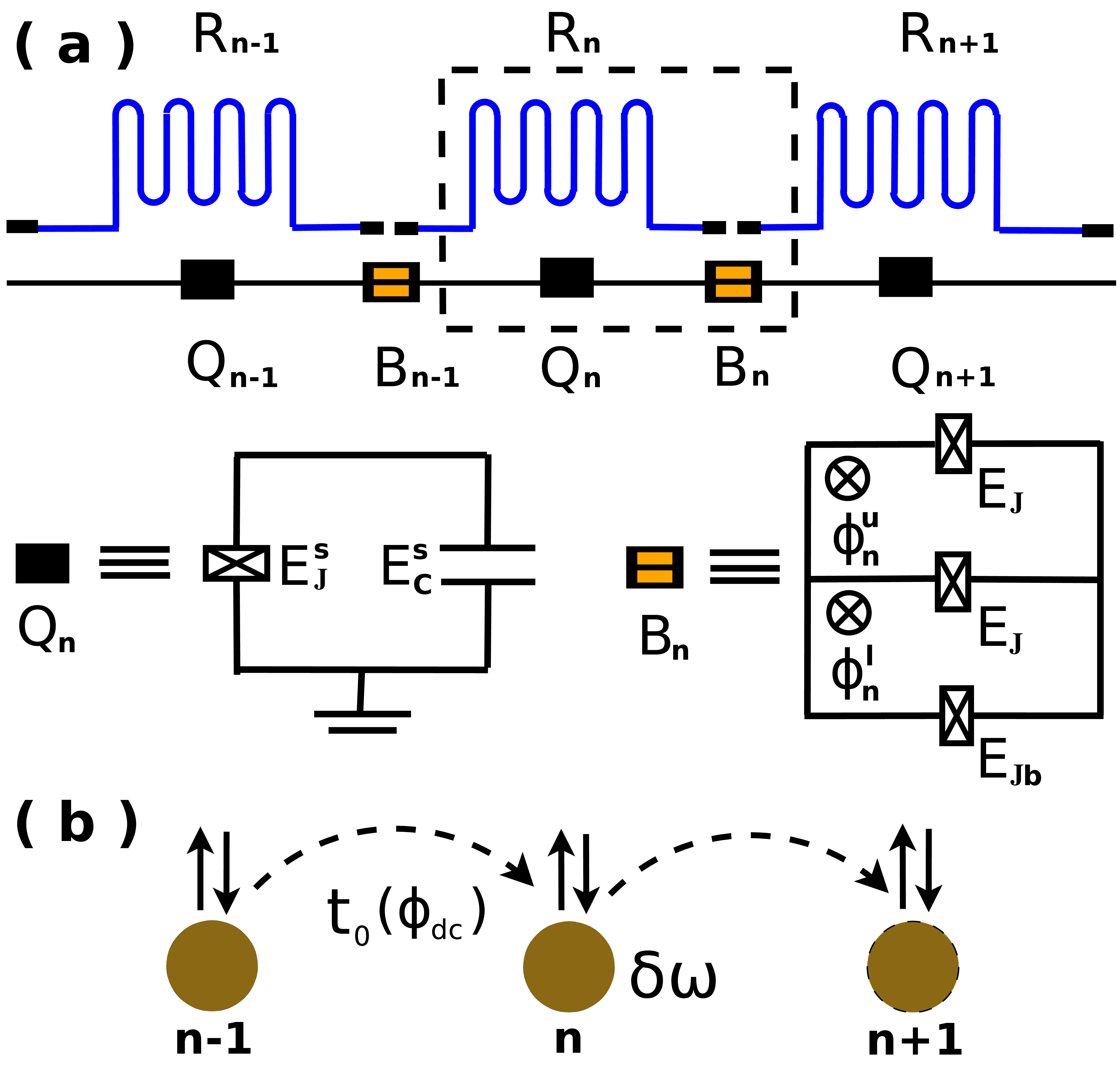}
\caption{\label{fig:circuit}(Color online)(a) Schematic of the qubit-resonator system, whose $n$-th repeating unit (indicated 
by the dashed rectangle) comprises SC qubit $Q_{n}$, resonator $R_{n}$, and coupler circuit $B_{n}$. The fluxes from the 
resonator modes $n$ and $n+1$ thread the upper loops of the coupler circuit $B_{n}$, effectively giving rise to an indirect 
inductive qubit-resonator coupling. (b) Pictorial illustration of the effective lattice model of the system, with the excitation 
hopping amplitude $t_{0}(\phi_{\textrm{dc}})$ and each lattice site hosting dispersionless bosons with the frequency 
$\delta\omega=\omega_c-\omega_0$.}
\end{figure}

\textit{Qubit-resonator system}.-- The analog simulator of the model under consideration
[see Fig.~\ref{fig:circuit}(a)] consists of SC qubits ($Q_{n}$) with the energy splitting $\varepsilon_{z}$, 
microwave resonators ($R_{n}$) with the photon frequency $\omega_{c}$, and coupler circuits ($B_{n}$)~\cite{Geller+:15}
which mediate both qubit-qubit and qubit-resonator interactions in this system. The simulator can be realized with 
transmons~\cite{Koch++:07} ($E^{s}_{J}/E^{s}_{C}\sim 100$, where $E^{s}_{C}$ and $E^{s}_{J}$ are the single-qubit 
charging- and Josephson energies) or gatemons~\cite{Larsen++:15} ($E^{s}_{J}/E^{s}_{C}\sim 25$). Its $n$-th repeating 
unit is described by the free Hamiltonian $H_{n}^{0}=(\varepsilon_{z}/2)\:\sigma_{n}^{z}+\hbar\omega_{c}\:b_{n}^{\dagger}b_{n}$, 
where the pseudospin-$1/2$ operators $\bm{\sigma}_n$ represent qubit $n$ and the bosonic operators ($b_{n},b_{n}^{\dagger}$) 
photons in the $n$-th resonator. 

The upper and lower loops of $B_{n}$ are threaded by magnetic fluxes $\phi_{n}^{u}$ and $\phi_{n}^{l}$, respectively 
[both are expressed in units of $\Phi_{0}/2\pi$, where $\Phi_{0}\equiv hc/(2e)$ is the flux quantum]. In particular, 
the upper loop is subject to ac-driving with the flux $\pi\cos(\omega_{0}t)$. The other contribution to $\phi_{n}^{u}$ 
originates from the modes of resonators $n$ and $n+1$ and is given by $\phi_{n,\textrm{res}}=\delta\theta[(b_{n+1}+
b_{n+1}^{\dagger})-(b_{n}+b_{n}^{\dagger})]$, where $\delta\theta=[2eA_{\textrm{eff}}/(\hbar d_{0}c)]\times(\hbar
\omega_{c}/C_{0})^{1/2}$, with $A_{\textrm{eff}}$ being the effective coupling area, $C_{0}$ the resonator capacitance, 
and $d_{0}$ the effective spacing in the resonator~\cite{OrlandoBook}. Therefore, unlike the much more common capacitive 
coupling~\cite{Mei+:13}, the qubit-resonator coupling in the system at hand is inductive~\cite{SCqubitReview:19}. Similarly, 
$\phi_{n}^{l}$ includes an ac contribution, given by $-(\pi/2)\cos(\omega_{0}t)$, and a dc part $\phi_{\textrm{dc}}$, 
the main experimental knob in this system.

The Josephson energy of $B_{n}$ is given by $H_{n}^{J}=-\sum_{i=1}^{3}E^{i}_{J}\cos\varphi_{n}^{i}$, with $\varphi_{n}^{i}$ 
being the respective phase drops on the three Josephson junctions within $B_{n}$ and $E^{i}_{J}$ their energies,
here chosen such that $E^{1}_{J}=E^{2}_{J}\equiv E_{J}$ and $E^{3}_{J}=E_{Jb}\neq E_{J}$. Using the 
flux-quantization rules~\cite{VoolDevoretReview:17}, the total Josephson energy $\sum_n\:H_{n}^{J}$ can be 
expressed in terms of the gauge-invariant phase variables $\varphi_{n}$ of SC islands of different qubits.
The latter enter this energy through terms of the type $\cos(\varphi_{n}-\varphi_{n+1})$, which in the regime of 
interest for transmons/gatemons ($E^{s}_{J}\gg E^{s}_{C}$) can be recast (up to an additive constant) as 
$\delta\varphi_{0}^{2}\left[\sigma_{n}^{+}\sigma_{n+1}^{-}+\sigma_{n}^{-}\sigma_{n+1}^{+}-(\sigma_{n}^{z}+
\sigma_{n+1}^{z})/2\right]$, where $\delta\varphi_{0}^{2}\equiv (2E^{s}_{C}/E^{s}_{J})^{1/2}$~\cite{Mei+:13}. 

Further analysis is carried out in the rotating frame of the drive, where $\delta\omega\equiv\omega_c-\omega_0$ is the 
effective boson frequency and the Josephson-coupling term becomes time-dependent. This time dependence can, however, be 
disregarded due to its rapidly-oscillating character, in line with the rotating-wave approximation (RWA). The remaining 
part of $H_{n}^{J}$ can succinctly be written as $\bar{H}_{n}^{J}=-\mathcal{E}_{n}^{J}(\phi_\textrm{dc},
\phi_{n,\textrm{res}})\:\cos(\varphi_{n}-\varphi_{n+1})$, where
\begin{equation}\label{eq:HJn_final}
\mathcal{E}_{n}^{J}=E_{Jb}\left(1+\cos\phi_{\textrm{dc}}\right)
-E_{J}J_{1}(\pi/2)\phi_{n,\textrm{res}}\:,
\end{equation}
and $J_m(x)$ are Bessel functions of the first kind whose presence in this expression stems from the use of the Jacobi-Anger 
expansion~\cite{AbramowitzStegunBOOK} in conjunction with the RWA. In what follows, without significant loss of generality 
$E_{Jb}$ is chosen to be given by $2E_{J}J_{0}(\pi/2)$. 

The above expression for $\cos(\varphi_{n}-\varphi_{n+1})$ in terms of the operators $\bm{\sigma}_n$ implies that the effective
interaction between adjacent qubits in this system is of $XY$ type. Through the flux $\phi_{n,\textrm{res}}$, the interaction 
strength acquires a dependence on the boson displacements $u_n \propto b_{n}+b_{n}^{\dagger}$
whose form is equivalent to that of the $XY$ spin-Peierls model~\cite{Caron+Moukouri:96}. The spinless-fermion -- boson coupling 
that results from this interaction via the JW transformation is nonlocal in nature, in contrast to other examples of such 
couplings in various solid-state systems~\cite{Palyi+:12,Souza+:19}.

\textit{Effective Hamiltonian and its ground states}.-- To show that the effective system Hamiltonian consists of 
contributions akin to $H_{0}$ and $H_{\textrm{e-b}}$ [cf. Eqs.~\eqref{H_0} and \eqref{H_eb}], one switches to the spinless-fermion 
representation using the Jordan-Wigner (JW) transformation. The latter reads $\sigma_{n}^{z} =  2c_{n}^{\dagger}c_{n}-1$, 
$\sigma_{n}^{+}=2c_{n}^{\dagger}\:e^{i\pi\sum_{l<n}c_{l}^{\dagger}c_{l}}$, where the sum in the last exponent defines the 
JW string~\cite{ColemanBOOK}. In this representation, the noninteracting part of the effective system Hamiltonian comprises 
the excitation-hopping- and free-photon terms. [Note that $c_{n}^{\dagger}c_{n}$ terms resulting from the $\sigma_{n}^{z}$ 
terms in $H_{n}^{0}$ and $\bar{H}_{n}^{J}$ are largely immaterial for further discussion as they only lead to a constant energy 
offset (band-center energy).] It assumes the form of $H_{0}$, with $\omega_{\textrm{b}}
\rightarrow\delta\omega$ and $t_{e}\rightarrow t_{0}(\phi_{\textrm{dc}})\equiv E_{Jb}\delta
\varphi_{0}^{2}\:\left(1+\cos\phi_{\textrm{dc}}\right)$, the latter being the effective $\phi_{\textrm{dc}}$-dependent 
hopping amplitude [cf. Fig.~\ref{fig:circuit}(b)]. At the same time, the interacting part adopts the form of $H_{\textrm{e-b}}$. 
In particular, via the JW transformation the Peierls-coupling term is obtained in a manner familiar from the $XY$ spin-Peierls 
model~\cite{Caron+Moukouri:96}, while the breathing-mode term originates from the $\sigma_{n}^{z}$ terms in the above 
expression for $\cos(\varphi_{n}-\varphi_{n+1})$.

The dimensionless coupling strength $g$ is determined by the system parameters through the relation 
$g\hbar\delta\omega=\delta\varphi_{0}^{2}\:E_{J}J_1(\pi/2)\delta\theta$, while the $\phi_{\textrm{dc}}$-dependent -- thus 
{\em in-situ} tunable -- effective coupling strength is given by
\begin{equation}\label{expr_lambda}
\lambda_{\textrm{e-b}}(\phi_{\textrm{dc}})=g\:\frac{J_1(\pi/2)\delta\theta}
{J_0(\pi/2)\left(1+\cos\phi_{\textrm{dc}}\right)} \:.
\end{equation}
For a typical resonator $\delta\theta\sim 3.5\times 10^{-3}$~\cite{Stojanovic+Salom:19}. Besides, for 
$\delta\omega$ it is pertinent to take $\delta\omega/2\pi=200 - 300$ MHz and also choose $E_{J}$ 
such that $\delta\varphi_{0}^{2}\:E_{J}/2\pi\hbar=100$ GHz. 

The ground-state energy of the system, expressed in units of $E_{\textrm{u}}\equiv\:10^{-3}\:\delta\varphi_{0}^{2}\:E_{J}$,
was evaluated through Lanczos-type exact diagonalization~\cite{CullumWilloughbyBook,Stojanovic+Salom:19} and 
illustrated (without the constant-energy contribution) in Fig.~\ref{fig:Energy}. 
For $\lambda_{\textrm{e-b}}\ge\lambda^{\textrm{c}}_{\textrm{e-b}}$ the system has a 
polaron-like ground state (strongly boson-dressed excitation), with its energy showing a rather weak dependence on $\lambda_{\textrm{e-b}}$. 
On the other hand, for $\lambda_{\textrm{e-b}}<\lambda^{\textrm{c}}_{\textrm{e-b}}$ the ground state corresponds to 
$|\Psi_{k=0}\rangle$, i.e., a bare excitation with $k=0$. Its energy $E_{\textrm{gs}}=-2t_{0}(\phi_{\textrm{dc}})$ 
is the minimum of a 1D cosine-shaped dispersion. The energy separation of this ground state from the first excited state exactly equals 
$\hbar\delta\omega$ for any $\phi_{\textrm{dc}}$ below the critical value. This is consistent with the fact that coupled e-b systems 
with dispersionless bosons invariably have one-boson continua separated from their ground states by the single-boson energy and 
in the weak-coupling regime typically feature only one bound state below those continua~\cite{Ku+Trugman:07}.

\textit{$W$ states and their preparation}.-- Bearing in mind that JW strings act trivially on $|0\rangle_{\textrm{e}}$, 
so that $c^{\dagger}_{n}|0\rangle_{\textrm{e}}\equiv S_{n}^{+}|0\rangle_{\textrm{e}}$ (where $\mathbf{S}_n\equiv\hbar
\bm{\sigma}_n/2$), it holds that  
\begin{equation}
c^{\dagger}_{k}|0\rangle_{\textrm{e}}=N^{-1/2}\sum_{n=1}^{N}\:e^{-ikn}\:S_{n}^{+}|0\rangle_{\textrm{e}} \:.
\end{equation}
The last equation is equivalent to $|\Psi_{k}\rangle=|W_N(k)\rangle\otimes|0\rangle_{\textrm{b}}$, where $|W_N(k)\rangle$ 
is a ``twisted'' $N$-qubit $W$ state. Thus, bare-excitation Bloch states coincide with generalized $W$ states, while in 
particular $|\Psi_{k=0}\rangle$ -- the ground state of the system at hand for $\lambda_{\textrm{e-b}}<\lambda^{\textrm{c}}_
{\textrm{e-b}}$ -- corresponds to the ordinary $N$-qubit $W$ state 
\begin{equation}
|W_N\rangle =\frac{1}{\sqrt{N}}\left(|10\ldots 0\rangle+
|01\ldots 0\rangle+\ldots+|00\ldots 1\rangle\right) \:.
\end{equation}

In the following, a microwave-pumping based protocol for the preparation of an $N$-qubit $W$ state is proposed assuming 
that the system is initially in the vacuum state $|0\rangle\equiv|0\rangle_{\textrm{e}}\otimes|0\rangle_{\textrm{ph}}$. 
The external driving required for this purpose is assumed to be represented by the operator
\begin{equation}\label{driveop}
\Omega_{q_d}(t)=\frac{\hbar\beta(t)}{\sqrt{N}}\sum_{n=1}^{N}\left
(\sigma_{n}^{+}e^{-iq_d n}+\sigma_{n}^{-}e^{iq_d n}\right) \:,
\end{equation}
where $\beta(t)$ describes its time dependence and the factors $e^{\pm iq_d n}$ account for the possibility that flip 
operations on different qubits are applied with a phase difference. It is important to stress that the general form of 
driving in Eq.~\eqref{driveop} allows one to prepare -- through different choices of $q_d$ and $\beta(t)$ -- various states 
of the proposed system, including its strongly boson-dressed ground states realized when $\phi_{\textrm{dc}}$ is above the 
critical value.

The transition matrix element of the operator $\Omega_{q_d}(t)$ between the initial state $|0\rangle$ and the target 
state $|\Psi_{k=0}\rangle\equiv|W_N\rangle\otimes |0\rangle_{\textrm{ph}}$ evaluates to $\hbar\beta(t)\:\delta_{q_d,k=0}$, 
which indicates that the preparation of this particular target state requires only a global driving field [i.e., $q_d=0$ 
in Eq.~\eqref{driveop}]. Thus, in contrast 
to some other schemes for $W$-state preparation~\cite{Gangat+:13}, the present one does not require a local qubit 
control~\cite{Heule++,Stojanovic:19}. By assuming that $\beta(t)=2\beta_{p}\cos(\omega_{d}t)$, where $\hbar\omega_{d}$ 
is the energy difference between the two relevant states, in the RWA these 
states are Rabi-coupled with the effective Rabi frequency $\beta_{p}$~\cite{Russ+:18,Zajac+:18}. Thus, starting from 
the state $\left|0\right\rangle$, the desired state $N$-qubit $W$ state will be prepared within a time interval of 
duration $\tau_{\textrm{prep}}=\pi\hbar/(2\beta_{p})$, which does not depend on $N$. 

Taking the pumping amplitude to be $\beta_{p}/(2\pi\hbar)=10$\:MHz, one finds $\tau_{\textrm{prep}}\approx 25$\:ns,
which is three orders of magnitude shorter than typical coherence times of SC qubits (e.g., for transmons 
$T_2\sim 20-100\:\mu$s~\cite{SCreviews:17}). Thus the proposed protocol should not be affected by a loss of coherence 
in the system. At the same time, the obtained $\tau_{\textrm{prep}}$ is sufficiently long that a leakage outside of the 
computational subspace of a single qubit can be neglected. Namely, due to the multilevel character of SC qubits, a finite 
anharmonicity $\alpha\equiv E_{12}-E_{01}$ (where $E_{ij}$ is the energy difference between qubit states $j$ and $i$) is 
required. In order to avoid such a leakage, the minimal pulse duration of $t_p\sim \hbar/|\alpha|$ is necessary. 
For transmons ($\alpha\sim -200$\:MHz), even a few-nanoseconds-long pulse is frequency selective enough that such a leakage 
is negligible~\cite{GirvinCQEDintro}. The obtained $\tau_{\textrm{prep}}\sim 25$\:ns suffices even in the case of gatemons, 
whose typical anharmonicity is by a factor of two smaller than that of transmons~\cite{Kringhoj+:18}.

Importantly, the large energy separation $\hbar\delta\omega$ between the target state and the lowest-lying excited 
state of the system ensures that the proposed $W$-state preparation will not be hampered by an inadvertent population of undesired 
states. For instance, for $\delta\omega/2\pi=200\:(300)$\:MHz this energy separation is equal to $2E_{\textrm{u}}\:(3E_{\textrm{u}})$, 
which represents a significant fraction of the energy difference between the initial and target states (cf. Fig.~\ref{fig:Energy}).

The proposed protocol is deterministic in nature and generates $W$-type entanglement of all the qubits in parallel. Moreover, in 
contrast to the typical situation in quantum-state control, where state-preparation times often scale unfavorably with the system 
size, here $\tau_{\textrm{prep}}$ does not depend on the system size at all. Finally, because they represent ground states of the 
system, multipartite $W$ states prepared by this protocol can be expected to be extremely robust.  

Besides allowing $W$-state preparation, the proposed system features an $XY$-type qubit-qubit interaction,  
which opens the possibility for a universal quantum computation~\cite{Schuch+Siewert:03,Abrams+:19}. Because the strength of this 
interaction depends dynamically on the boson degrees of freedom (photons), this system bears a formal similarity to certain 
trapped-ion systems in which the role of bosons is played by collective motional modes (phonons)~\cite{Wall+:17}. Compared to its 
trapped-ion counterparts, this system has an added advantage that it merely involves dispersionless bosons of one single frequency,
which circumvents the spectral crowding problem resulting from the quasicontinuous character of phonon spectra in large trapped-ion 
chains~\cite{Landsman+:19}. 

\textit{Robustness to losses and feasibility}.-- It is pertinent to briefly address the robustness of the system at hand to possible 
deleterious effects of losses. To this end, it is worthwhile to first note that qubit-state flips and displacements of the resonator 
modes are the two leading sources of decoherence in this system. In addition to the very long $T_2$ times of transmon 
(gatemon) qubits, the damping time of microwave photons in coplanar waveguide resonators can reach the same order of magnitude as $T_2$, 
with the corresponding quality factor being larger than $10^7$~\cite{Wang+:19}. Besides, the relevant excitation- and photon energy scales 
in this system ($\delta\omega$, $g\delta\omega$, $t_0/\hbar$), expressed in frequency units, are all of the order of several 
$2\pi\times 100$\:MHz. Thus, they far exceed the decoherence rates whose state-of-the-art values in this type of systems are $\gamma\sim
0.01$\:MHz~\cite{SCreviews:17}. Finally, in this system thermal excitations -- which at temperatures typical for such SC-qubit setups 
($T\sim 100$\:mK) have characteristic energies of a few GHz -- can be safely neglected. Therefore, the loss mechanisms do not pose 
obstacles to realizing the proposed system.

\textit{Conclusions}.-- The present paper proposes a scheme for a fast, deterministic creation of a large-scale  
$W$-type entanglement~\cite{Froewis+:18} in a system of inductively-coupled superconducting qubits and microwave 
resonators. The mechanism behind this scalable entanglement resource -- which allows one to engineer $W$ states with the preparation 
times independent of the system size -- stems from the unconventional ground-state properties of a one-dimensional model 
describing a nonlocal coupling of a spinless fermion to zero-dimensional bosons. The feasibility and robustness of the underlying 
state-preparation protocol -- which only requires a global driving field -- is demonstrated with realistic system parameters. 

This study can be viewed as being complementary to that of Ref.~\onlinecite{Gangat+:13}, where the preparation of $W$ states of 
photons -- rather than qubits -- was proposed. The common denominator of these two proposals is that they both rely on superconducting 
systems and an {\em in-situ} tunability of a hopping amplitude, albeit being based on completely different physical mechanisms. 
These schemes are far more scalable than the conventional ones in which resonator-mediated qubit-qubit interactions are utilized 
to controllably entangle multiple qubits; such an approach was recently used to prepare a GHZ state of $10$ superconducting 
qubits~\cite{Song+:17} -- the largest entanglement demonstrated so far in solid-state architectures. Thus, the need to demonstrate 
the envisioned $W$-state preparation is compelling.
\begin{acknowledgments}
\textit{Acknowledgments}.-- This research was supported by the Deutsche Forschungsgemeinschaft (DFG) -- SFB 1119 -- 236615297.
\end{acknowledgments}

\end{document}